# Comparing Classical Pathways and Modern Networks: Towards the Development of an Edge Ontology

Long J. Lu[1], Andrea Sboner[1], Yuanpeng J. Huang[4], Hao Xin Lu[1], Tara A. Gianoulis[1], Kevin Y. Yip[2], Philip M. Kim[1], and Gaetano T. Montelione[4,5], Mark B. Gerstein[1, 2, 3*]

1. Department of Molecular Biophysics and Biochemistry, Yale University, 266 Whitney Avenue, New Haven, CT 06520

2. Department of Computer Science, Yale University, 51 Prospect Street, New Haven, CT 06511

3. Program in Computational Biology and Bioinformatics, Yale University, 266 Whitney Avenue, New Haven, CT 06520

4. Department of Molecular Biology and Biochemistry, Center for Advanced Biotechnology and Medicine, Rutgers University, Piscataway, NJ 08854

5. Department of Biochemistry, Robert Wood Johnson Medical School, UMDNJ, Piscataway, NJ 08854.

* Corresponding author: Gerstein, M.B. (Mark.Gerstein@yale.edu)

## Abstract

Pathways are integral to systems biology. Their classical representation has proven useful but is inconsistent in the meaning assigned to each arrow (or edge) and inadvertently implies the isolation of one pathway from another. Conversely, modern high-throughput experiments give rise to standardized networks facilitating topological calculations. Combining these perspectives, we can embed classical pathways within large-scale networks and thus demonstrate the crosstalk between them. As more diverse types of high-throughput data become available, we can effectively merge both perspectives, embedding pathways simultaneously in multiple networks. However, the original problem still remains – the current edge representation is inadequate to accurately convey all the information in pathways. Therefore, we suggest that a standardized, well-defined, edge ontology is necessary and propose a prototype here, as a starting point for reaching this goal.

## Uniting classical pathways and modern networks

In biology, a pathway refers to a sequence of reactions, usually controlled and catalyzed by enzymes, by which one organic substance is converted to another. Biological pathways are an important component of systems biology. The classical representation of these pathways provides varied, mechanistic associations between many proteins. Conversely, modern high-throughput experiments and large-scale databases have given rise to standardized networks that provide a somewhat different perspective on pathways. Combining and comparing these perspectives, classical biochemical pathways can be embedded into large-scale networks. This reveals a number of problematic issues with classical pathways: First, the same pathways documented in different databases are often inconsistent in components included and their exact symbolic representation (e.g., what each of the arrows mean). Second, pathways are isolated from one another in classical representations, de-emphasizing crosstalk. In contrast, embedded pathways offer completely uniform representations and relate network statistics, such as average degree or diameter, consistently to pathways. However, they are more limited in the richness of the mechanistic biochemistry that they can convey. As more diverse types of high-throughput data become available, it will be possible to embed classical pathways simultaneously in many large-scale networks, effectively merging both approaches. To accomplish this, a precise edge (or arrow) ontology needs to be defined. For illustrative purposes, we propose a prototype of ontology which provides an unambiguous representation of the edges connecting biomolecules and also describes higher-level relationships amongst these edges. We then demonstrate the usefulness of the simple edge ontology on four diverse types of pathways. We do not intend to provide a complete ontology here, but rather to stimulate people working in this field to continue building upon existing knowledge until a complete ontology is achieved.

## Pathway databases and limitations

During the last decade, an increasing number of pathway databases have been established in order to document the ever-expanding knowledge concerning established pathways. Some of these pathway databases are organism specific. For example, EcoCyc[1] describes the genome and the biochemical machinery of *E. coli* K12 MG1655. A few other pathway databases focus on a specific type of disease or disorder, such as The Cancer Cell Map (http://cancer.cellmap.org) or GOLD.db[2]. The majority of these pathway databases cover a certain functional area that occurs in multiple organisms. Furthermore, such databases can often be roughly divided into three categories: those containing metabolic pathways (KEGG[3], WIT[4], BioCyc[5], MetaCyc[6], and GenMAPP[7]); those containing signal transduction (signaling) pathway (BioCarta (http://biocarta.com), STKE (http://stke.org), Pathways Knowledge Base (http://ingenuity.com), and Reactome[8]); and those containing both (KEGG, BioCarta and Reactome). Excellent recent reviews on these pathway databases can be found elsewhere [9, 10].

Although the above-mentioned databases provide valuable resources for studying the associations between proteins, they are hampered by several limitations. First of all, the same pathways documented in different databases are often inconsistent. In many cases, a pathway is described by including a few core components first. The decision of whether to include additional components in the given pathway is usually empirically determined, based on the expert curators' knowledge and experience. Therefore, the boundary of a pathway is usually vague. The consequence is that the number of components in the same pathway in different databases varies greatly (Table S1 and Figure 4a).

Second, these pathways are isolated in classical representations. This is the consequence of the traditional reductionist approaches to molecular biology, whereby genes and pathways are investigated as isolated entities. However, from the perspective of modern systems biology, the interactions between biological pathways must be studied in order to understand how biological systems function. On the systems level, the crosstalk between pathways appears to be particularly important but lack significant study. Although there have been efforts to integrate them, such as in the Boehringer Mannheim Biochemical Pathways wall chart, many aspects of the relationships between pathways have yet to be systematically identified and incorporated.

Third, the classical representations of pathways use symbols that lack a precise definition. The same symbol is often used to represent a variety of functions. For example, arrows are used to represent direct interactions in some circumstances, but in others, they are also used to represent translocation to a different sub-cellular compartment. Although this might not cause problems for laboratories focusing on individual pathways, these notations must be precisely defined in order to perform analyses on pathways on a larger scale. A structured vocabulary or ontology of these symbols should be developed to ameliorate this problem.

**Recent advent of network biology**

A particularly novel concept in the post-genomic era is the idea that a living cell can be viewed as a complex network of biomolecules. Indeed a biomolecular network can now be rendered as a collection of nodes and edges. Nodes represent biomolecules, such as proteins, genes and metabolites, while edges represent the types of associations between two nodes, such as physical interactions and co-expression of mRNAs. The combined functions and interactions between these networks constitute the behavior of the cell. Mapping and understanding biomolecular networks represents the first step towards modeling how a cell actually operates.

As a result of recent genome-wide high-throughput (HTP) experiments, including large-scale yeast two-hybrid screens and microarray experiments, many types of networks have been mapped, including protein-protein interaction (PPI), expression, regulatory, metabolic and signaling networks. For example, protein-protein interaction networks have been experimentally determined in *Saccharomyces cerevisiae*[11-15], *Caenorhabditis elegans*[16], *Drosophila melonogaster*[17], *Homo sapiens*[18, 19], *Plasmodium falciparum*[20] and *Helicobacter pylori*[21]. The availability of such well-mapped networks has allowed us to compare and contrast them in terms of global and local topology, as well as, to relate the structural properties of these networks to protein properties, such as function and essentiality.

Topological analysis of networks provides quantitative insight into their basic organization. Different network statistics have been designed to capture the characteristics of network topology (Table S2). Despite the seemingly vast differences between biomolecular networks, they are found to share common features with respect to network topology. Barabási *et al.* [22] proposed a “scale-free” model in which the degree distribution in many large networks follows a power-law distribution [ $P(k) \approx k^{-r}$ ]. What is remarkable about this distribution is that while most of the nodes within these networks have very few links, a few of these nodes, classified as hubs, are exceptionally well-connected. Concurrently, Watts and Strogatz[23] found that many networks also have a “small-world” property, meaning they are defined as being both highly clustered and containing small characteristic path lengths.

Network analysis has provided quantitative new insights into protein properties, cellular dynamics and other biological problems. For example, research has shown that hubs in a network are more likely to be essential proteins, as well as prompted debate about whether hubs tend to evolve slower [24-26]. Furthermore, different motifs have been implicated in different stages of dynamic transitions of a network[27].

**Comparisons between classical and embedded pathways**

One can construct large-scale networks using different types of data from HTP experiments: protein-protein interaction networks from yeast two-hybrid screens, and co-expression networks from micro-array experiments provide apt examples of this. For each classical pathway, we can extract the corresponding sub-network from the entire network by mapping the core components in this classical pathway onto the network of biomolecules. From a network point of view, this mapping can also be viewed as embedding pathway components into the network. To differentiate from the classical pathways, we refer to these sub-networks as embedded pathways (Figure 1). The core components of a classical or embedded pathway are defined as the biomolecules in the KEGG pathway diagram. KEGG is used in this article because of its high quality among pathway databases, as pointed out by Wittig *et al.* [28].

We will use the Notch pathway as an example to illustrate our findings because of its elegance and simplicity. The Notch signaling pathway is a highly conserved pathway for cell-cell communication. It is involved in the regulation of cellular differentiation and proliferation. As shown in Figure 1, we constructed core and extended embedded pathways by collecting the 22 core protein components listed in KEGG and mapping them onto the large-scale PPI network deposited in the Human Protein Reference Database (HPRD)[29]. The HPRD interactions are manually curated by expert biologists to reduce errors.

Comparisons between classical and core embedded Notch pathways reveal a number of differences. First, the classical pathway contains directed and undirected edges (Figure 2a). Directed edges often represent activations, such as the edge between Delta and Notch. They also represent translocation to a different cellular compartment, for example, the edge between Notch and NICD. Undirected edges often represent an interaction between two components, such as the edge between CSL and SKIP. In contrast, the edges between components in the embedded pathway are uniform (Figure 2b). In this case, they are protein-protein interactions. Although the edge representation in the core embedded pathway is more consistent, it loses information encoded in classical pathways.

Second, although most of the edges are common between both representations, some edges appear only in one representation. The core embedded pathway also reveals 12 new interactions that are not found in the KEGG classical pathway. Conversely, there are two edges that exist in KEGG, but are not present in the core embedded pathway. They are between Notch and DVL and between Notch and TACE suggesting either that the protein-protein interaction map is incomplete or that these interactions take place through an intermediate (Figure 2a, b).

Compared to the classical pathway, the core embedded pathway has two advantages: First, the core embedded pathway is able to suggest which isoform is responsible for an interaction. For example, in the interaction between Notch and Numb, the embedded pathway identifies that Notch1 (Entrez ID: 4851) but not the other three isoforms interacts with Numb (Figure 2c). In contrast, the current version of the classical pathway collapses multiple protein isoforms into one single node.

Second, extended embedded pathways can systematically suggest new components involved in classical pathways. The extended embedded Notch pathway identifies 218 new proteins that are potentially involved in the Notch pathway by extracting the immediate interacting partners of these core components (Figure 2d). It is increasingly evident that the Notch pathway is subject to a wide array of regulatory influences, from those that affect ligand-receptor interactions to those that govern the choice of Notch target genes [30, 31].

For example, the classical Notch pathway in KEGG shows that Dishevelled (DVL) inhibits Notch. In the HTP networks, we find that Notch and DVL do not interact directly, but through an intermediate protein between them, namely, glycogen synthase kinase 3β (GSK-3β). DVL and GSK-3β are known to be involved in the Wnt pathway. The interaction of Wnt with the Frizzled receptor results in the inhibition of GSK-3β, by means of DVL[32]. The relationship between GSK-3β and Notch has been found by Espinosa *et al.* [33]. Specifically, they report that GSK-3β is able to phosphorylate Notch2 protein both *in vitro* and *in vivo*. Their paper suggests that GSK-3β may be partially accounted for by crosstalk between Wnt and Notch pathways.

Despite the above-mentioned advantages, the embedded pathway suffers significant information loss by restricting the edges to describing physical interaction. One way to circumvent this problem would be to overlay additional types of large-scale data onto the network by defining different types of edges. For example, it has been found that Notch down-regulates PSEN. This interaction is particularly interesting because PSEN is a component of the γ-secretase complex which cleaves Notch's intracellular domain, triggering the rest of the pathway. By laying the regulatory network on top of the protein-protein interactions, this feedback loop is highlighted [34].

**Relating network properties in embedded pathways**

Because of the heterogeneity of the edges and the incomplete nature of classical pathways, it is difficult to relate the mathematical quantities of modern network biology to these pathways. However, the same task becomes straightforward when applied to the embedded pathways created by mapping the core components of classical pathways onto large-scale networks. We provide an illustrative example in the supplemental materials showing how the topological quantities in modern network biology can lead to new insights into biochemical pathways. We found that signaling pathways from metabolic pathways have significantly different network topologies (Tables S3 and Figure S1). This difference has allowed us to successfully differentiate signaling pathways from metabolic pathways.

It is also interesting to note that the topological quantities between two of the same type pathways (signaling or metabolic) can be different even when they contain a similar number of core components, as illustrated by Notch vs. Hedgehog extended embedded pathways in Figure 3, Do these differences reveal anything about the underlying mechanisms of Notch and Hedgehog? First, we must consider it is possible that these differences are merely artifacts; that is, the protein interaction network is incomplete, and as the map expands, such differences will disappear. While certainly plausible, an alternative explanation is that these differences reflect real biological differences. This may be explained by the different regulatory mechanisms used by the two pathways. For example, Notch pathway might be subject to a larger degree of regulatory influences. Indeed, while our analysis suggests that while the core components of Notch embedded pathway have about 21 interacting partners on average, those of Hedgehog embedded pathway have just about 7 (Figure 3).

In addition to global inferences, topological measures can be used to identify nodes of particular importance or function. For instance, hubs (nodes of high degree) in regulatory networks correspond to master regulators [35]. Conversely, bottlenecks (nodes of high betweenness) often correspond to nodes that act as important information conduits, particularly in directed networks such as metabolic networks (in which metabolites flow between nodes) or signaling networks (in which information in the form of activations flows between nodes) [36]. Furthermore, these nodes most likely act as connectors between pathways, thereby mediating crosstalk (see below). Likewise, one can identify essential regulators by searching for composite hubs (i.e., nodes that have a high degree both in the metabolic and signaling network [37]). One can imagine using a combination of different statistics across many different networks in order to identify a number of key nodes. As our knowledge increasingly covers the different types of networks, an unambiguous and rigorous definition of protein function will eventually emerge from a combination of topological measures and network position [38]. That is, the importance of a protein is not only defined by its classical biochemical function, but also its position in the network. For example, hubs in Notch pathway include the HDACs, CREBBP, EP300, and DVL2; all of these nodes play a critical role in the regulation of the pathway. Furthermore, AXIN1 is a bottleneck in both the Hedgehog and the Wnt pathways, suggesting its importance for the information flow both within and between these pathways.

**Examining crosstalk between embedded pathways**

In living organisms, pathways are not isolated entities. From a systems biology perspective, pathways are linked together through crosstalk to perform biological functions as a system. In biology, the term crosstalk refers to the phenomenon that signal components in signal transduction can be shared between different signaling pathways, and responses to a signal-inducing condition (e.g., stress) can activate multiple responses in the cell or organism. This crosstalk can be exemplified by a particular protein, Protein Kinase C (PKC), which is shared by MAPK, Calcium, Phosphatidylinositol, Wnt, and VEGF signaling pathways. However, since classical pathways only contain core components, they are insufficient to study crosstalk. This is evidenced by the low overlaps between classical pathways, which serve as an indicator of the extent of crosstalk between them (Figure 4 and Table S4). In contrast, embedded pathways provide an excellent platform to examine crosstalk, since components of pathways are essentially embedded within a bigger network which allows systematic identification of overlapping and linking components.

Figure 4 and Table S4 show the overlaps between embedded pathways that correspond to signaling pathways in humans. We examine the overlaps between core, as well as, extended embedded pathways. The larger the overlap between embedded pathways, the more crosstalk takes place between the two pathways. Although most of the core embedded pathways do not overlap significantly, the corresponding extended embedded pathways often show a significant increase in overlap (Table S4). These results suggest that a large number of proteins exist as liaison components between pathways, and all of the signaling pathways can be connected with one degree of separation at most. A careful examination of these intermediate proteins may be useful in unraveling the mechanisms by which different pathways are related to each other.

**Developing a simple version of edge ontology for pathways**
As we mentioned before, classical pathway representation is often ambiguous, using the same symbol to represent different functions. In the post-genomic era, this problem is further confounded by the emergence of various types of high-throughput data. Different types of high-throughput data reveal different relationships between pathway components. In addition to protein-protein interaction networks, the core components of a pathway can also be mapped onto other types of networks, such as gene expression and regulatory networks. Simple edges and arrows that are traditionally used in the classical pathway representation may not be sufficient to meet the challenges of integrating these heterogeneous datasets. In order to perform large-scale mining of pathways, a precise edge or arrow ontology must be developed to represent different types of relationships between pathway components.

To make things more complicated, a large number of pathway databases are currently available [9, 39]. Unfortunately, they typically do not share data models, file formats or access method. To foster sharing of these different information sources, several eXtensible Markup Language (XML) exchange formats have been developed. System Biology Markup Language (SBML) [40] and CellML [41] focus mainly on quantitatively simulating concentrations of pathway components. The Proteomic Standards Initiative’s Molecular Interaction (PSI-MI) [42] is an exchange format for molecular interaction, and the Biological Pathway Exchange (BioPAX) [43] is a more general format used to describe biological pathways.

Much effort has been devoted towards developing consistent representation of pathways; however, most of these efforts focus on enumerating diverse types of edges. BioPAX has been developing an ontology of interactions that reveals relationships between edges. In order to perform large-scale mining of pathways, making explicit the relationships between edges is an important step for elucidating the transitions and reactions between molecules. A precise edge or arrow ontology may also help improve pathway representation by highlighting both different types of relationships between pathway components and our level of knowledge in that relationship.

As an example, phosphorylation, unbiquitination, glycosylation and methylation can all be viewed as types of reactions by transferring “tags” to target proteins. Thus an ontology of edges which not only enumerates different types of edges but also classifies the edges into groups should be developed to capture this information. This explicit hierarchy of relationships can be exploited to enable accurate computational analysis without losing the expressiveness of the classical representation. For example, some pathway interactions can be represented by specific symbols, such as serine phosphorylation. However, translation to a more general interaction, such as “tagging” that leads to activation, can be employed to perform high-level analysis of the pathway or to compare multiple pathways represented at a different level of specification.

Here, we propose a simple version of edge ontology to illustrate how we could deal with this issue in the future. This ontology provides both an unambiguous definition of the interactions and defines a hierarchy of those interactions. In particular, the hierarchy of interactions may be useful to obtain multiple views of a pathway, from a general one to a more specific one. It also contains symbols that may help the graphical representation of interactions. Please note this ontology is far from complete. We foresee that the formidable goal of constructing a complete ontology for pathways would take multiple groups many years to achieve. However, this simple ontology could be used as a starting point, upon which we hope a complete ontology can be built. We also realize that a consistent representation of nodes is equally important; however, this problem can be largely solved by using Gene Ontology[44]. The Gene Ontology (GO) provides a controlled vocabulary to describe gene and gene product attributes in any organism, and hence could be used as a rough node ontology. Below, we will only focus on edge ontology.

Table 1 shows the edge ontology. We use different shapes, symbols, and colors to represent diverse types of interactions between pathway components. We also define a simple hierarchy of interactions from general ones to more specific ones. The first level divides directed from undirected interactions whereas, the second level highlights the main mechanisms of interaction, which are in turn defined in more detail in the third level. The fourth level further specifies some of the interaction types. The edges in the second level have different shapes, while in the third level they are represented by different colors. Further specifications can be defined by adding annotations on the edge, such as those in the fourth level. Nearly all the edges connect two components of the pathways, such as proteins and molecules, except that the “catalysis” edge connects a pathway component to a “chemical reaction” edge. This allows us to properly describe metabolic pathways which typically display a sequence of chemical reactions in which enzymes take part.

As an example, a black arrow is used to indicate a “tagging” interaction, meaning an interaction that binds a molecule to a pathway component. If we know the type of interaction in more detail, different colors can be used to describe different “tagging” mechanisms: red for phosphorylation, blue for ubiquitination, and so on. In many cases, however, the tagging mechanism activates proteins; to highlight that some tagging mechanisms may inhibit proteins, a solid vertical line is used instead of the arrow shape. To further specify the relationship with more details, an annotation on the edge can be used, such as serine phosphorylation and N-linked glycosylation. It is worth noting that symbols from different levels can be used concurrently in the same pathway. This may be used as a way to emphasize our level of understanding of the interaction.

For illustration purposes only, we provide four examples of the application of this edge ontology. We consider the Notch pathway, the Citric acid cycle, the JAK-STAT signaling pathway and the Caspase cascade pathway. We redraw the classical pathways according to our new edge ontology in Figure 5.

## Concluding remarks

While classical representations of biochemical pathways can provide an in-depth view of isolated sets of genes, the network approach is capable of analyzing pathways on three different levels: whole system (crosstalk), whole network, and individual nodes. While embedding pathways to large-scale protein-protein interaction networks allows us to easily compare properties across, between, and within pathways, we also experience significant information loss. One way to circumvent this problem would be to overlay additional types of high-throughput data onto the pathway by defining different types of edges.

In order to properly analyze this type of multilayered network, a precise edge ontology must be defined. The edge ontology should provide an unambiguous representation of the relationships between biomolecules, as well as reveal relationships between edges. However, even a well-defined edge ontology still suffers the limitation of lacking explicit temporal information. Properly incorporating explicit temporal information will be the next grand challenge in the representation of pathways.

## ACKNOWLEDGMENTS

This work is supported by an NIH grant to MBG. We would like to thank Dr. Ashish Agarwal and Mr. Emmett Sprecher for valuable comments on improving this manuscript.

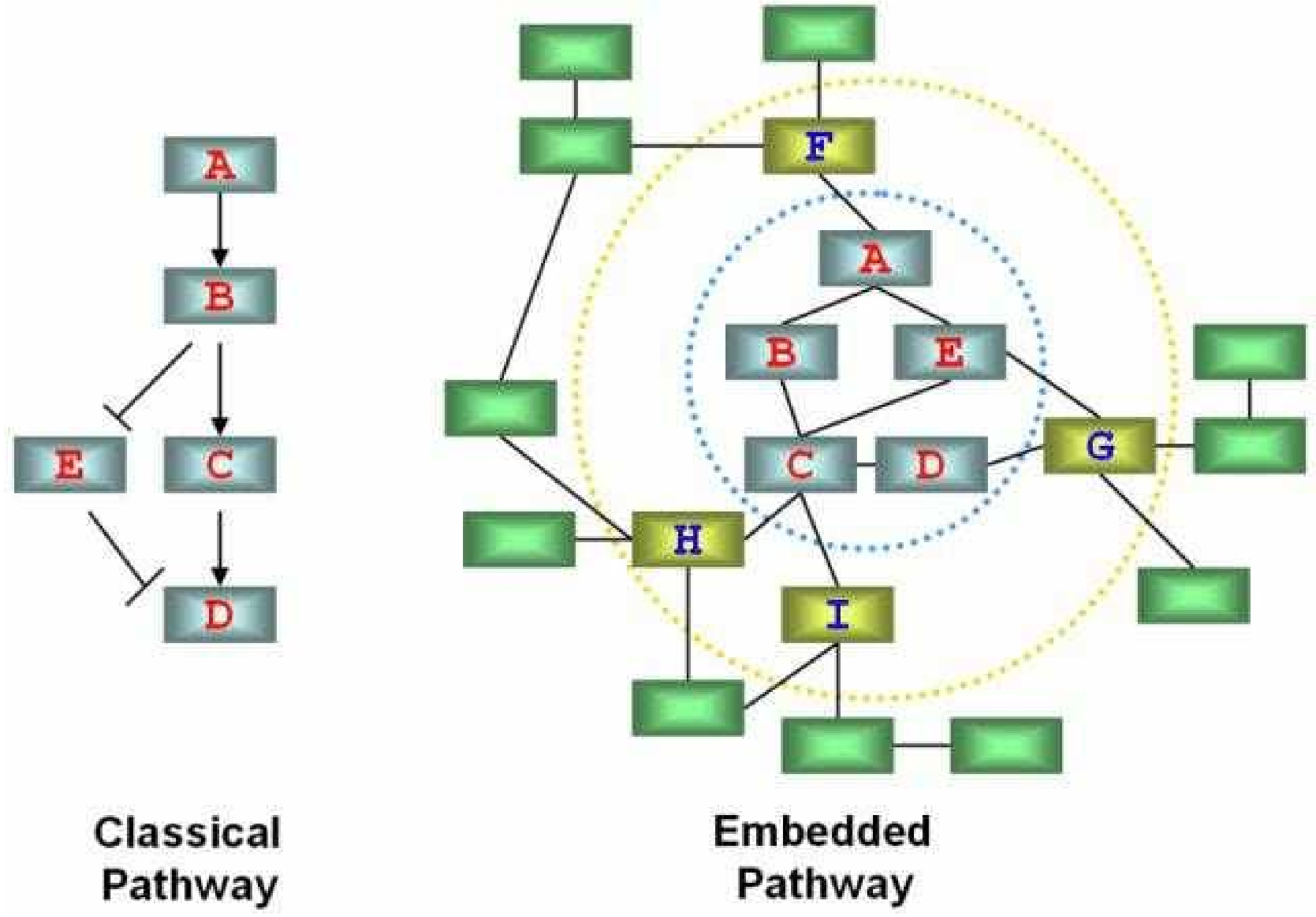

Figure 1 - Illustration of a classical and an embedded pathway.
A core embedded pathway contains the core components, as well as the edges linking them together (light blue nodes in Figure 1). The extended embedded pathway also contains the immediate nodes that are linked to the core components (light blue and yellow nodes). Comparisons on the classical and embedded pathways reveal interesting results.

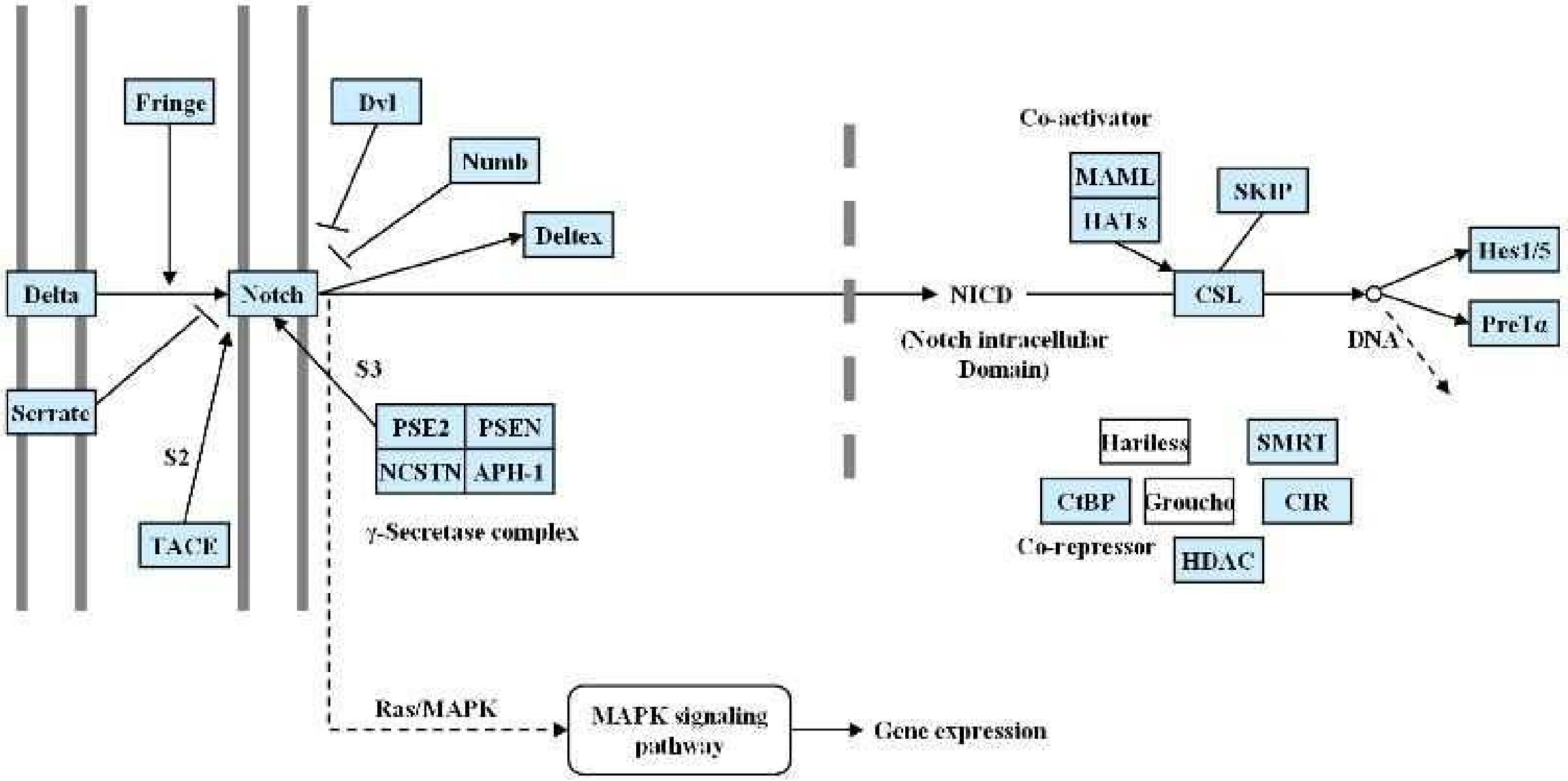

**Figure 2a Notch signaling pathway as illustrated in KEGG.**

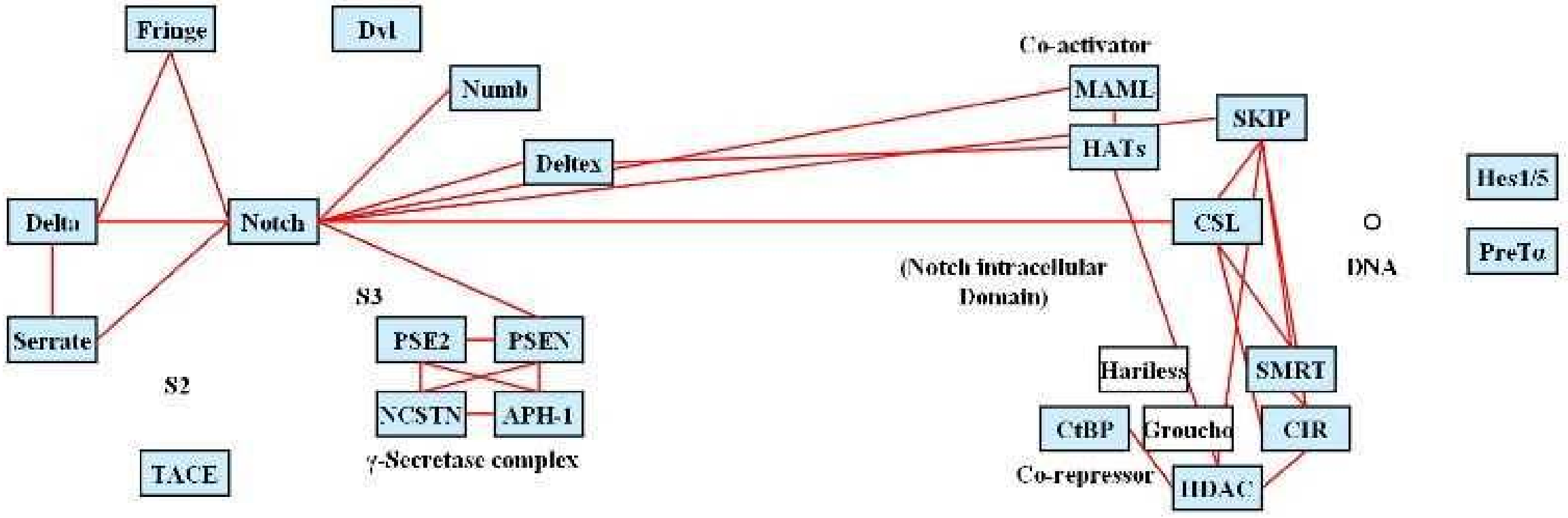

**Figure 2b Notch pathway mapped onto interaction networks. The edges represent protein-protein interactions.**

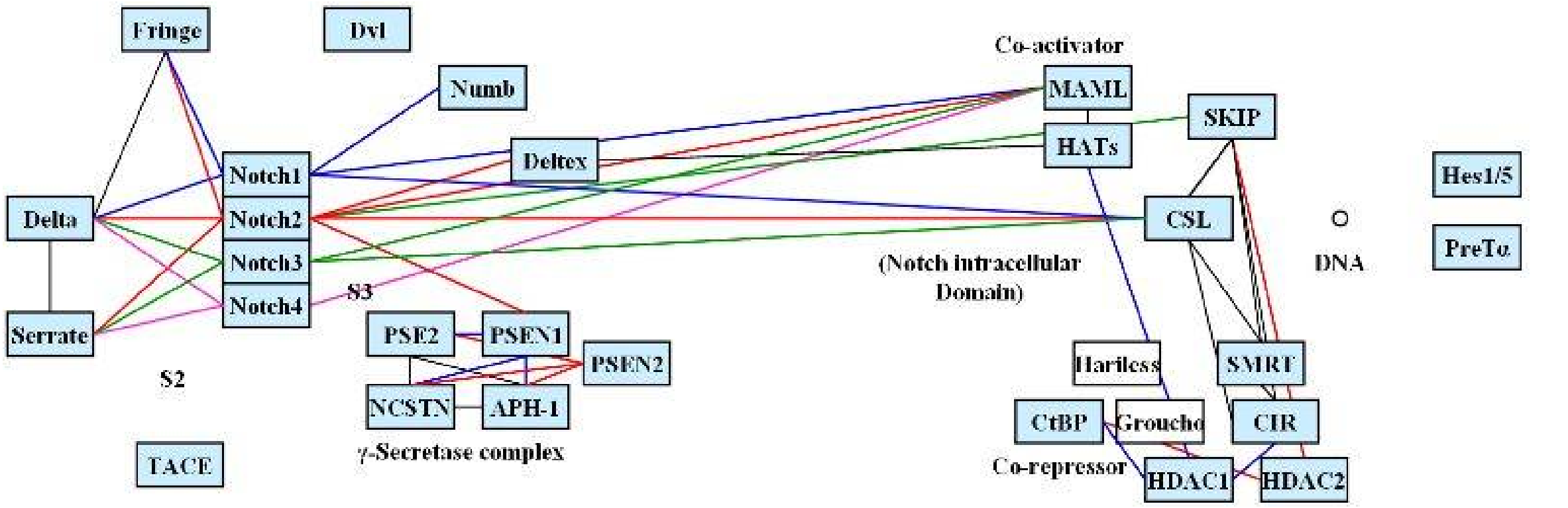

**Figure 2c Multiple nodes in the embedded Notch pathway. Different colors are used to differentiate the interactions between isoforms.**

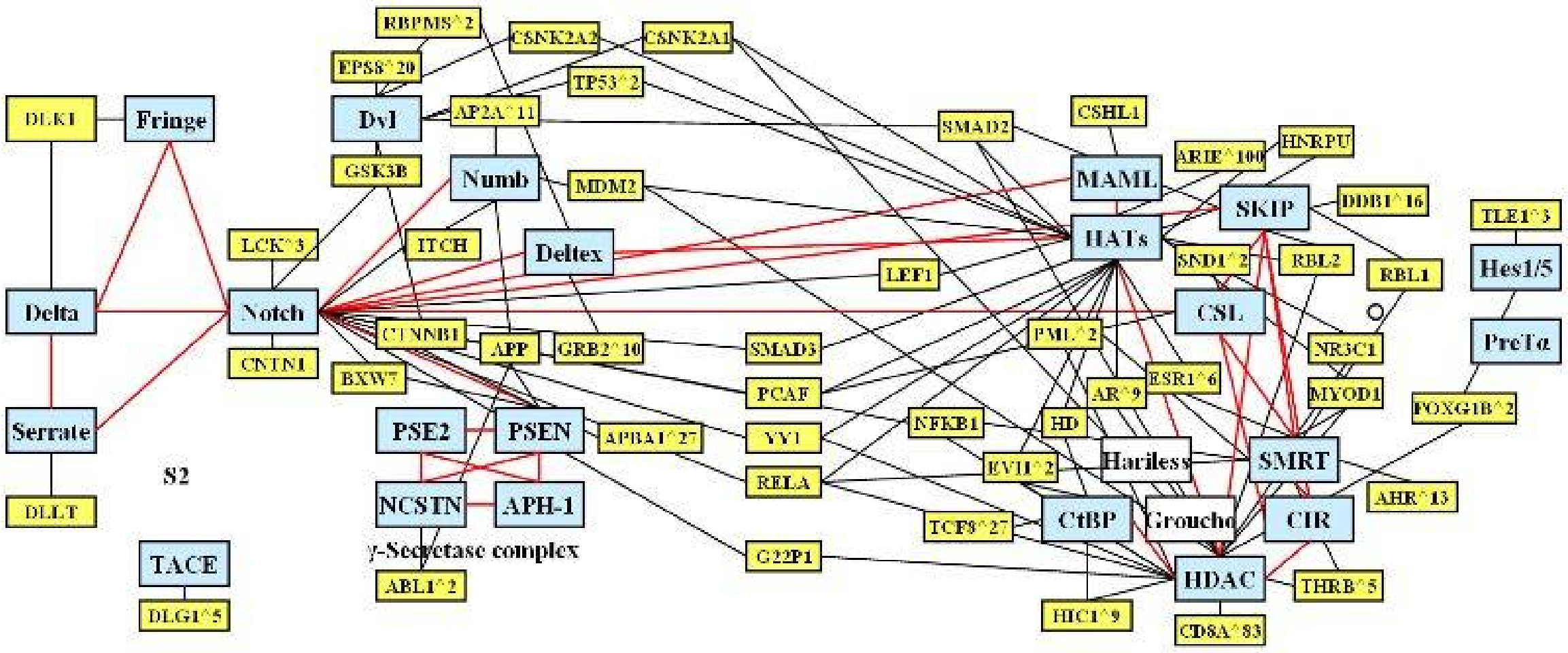

**Figure 2d First neighbors of Notch pathway proteins. Light blue nodes represent the core components of Notch pathway, and yellow nodes represent their interacting proteins. Red lines represent interactions between core components.**

A)

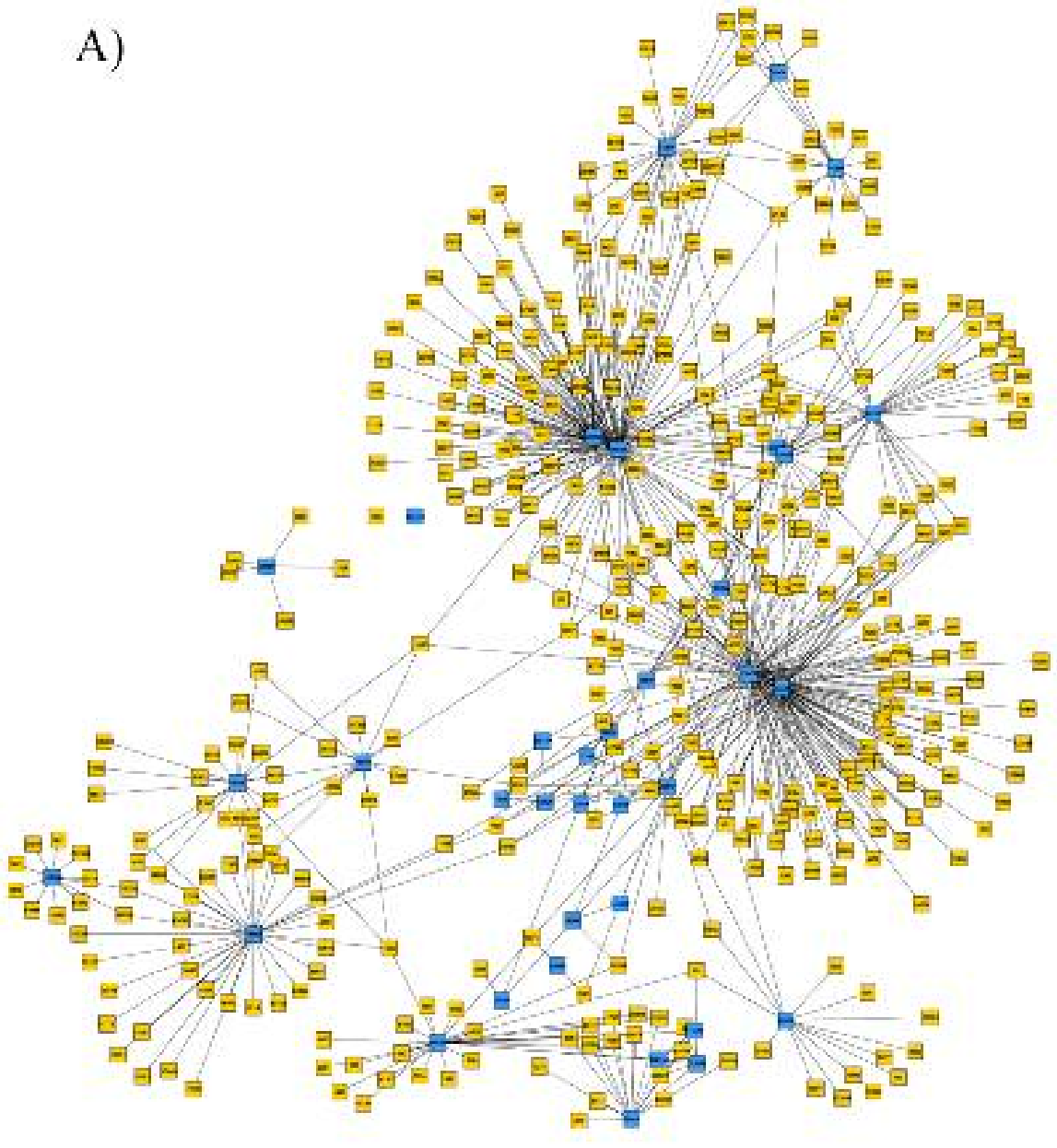

B)

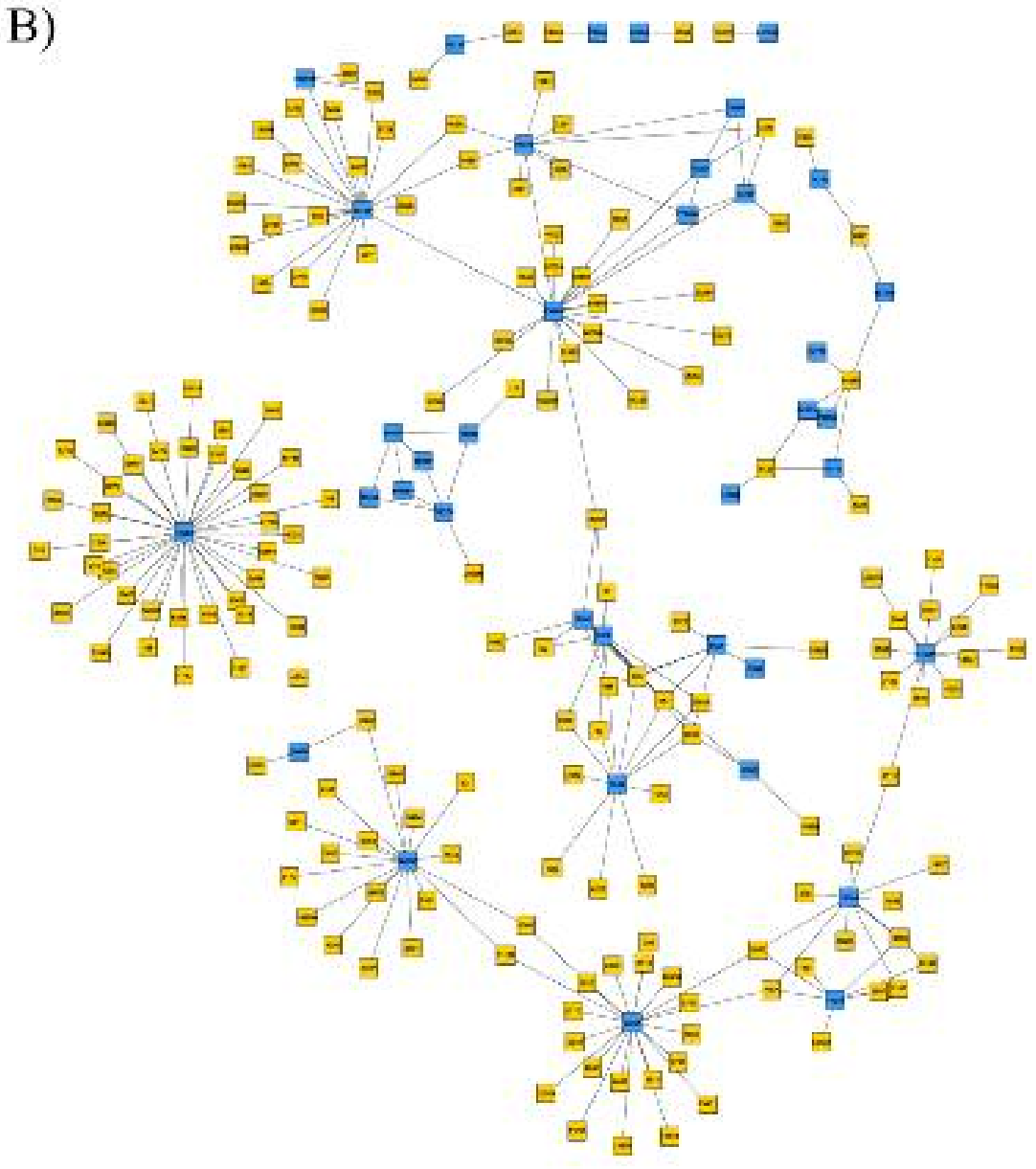

C)

| Pathway | Number of Nodes | Number of Edges | Connected Components | Number of Core Nodes | Number of Extended Node | Average Degree (K) | Clustering Coefficient (C) | Eccentricity (E) | Between-ness (B) |
|---|---|---|---|---|---|---|---|---|---|
| Notch | 459 | 683 | 3 | 36 | 423 | 20.47 | 0.17 | 6.64 | 6202.85 |
| Hedge-hog | 212 | 238 | 9 | 38 | 174 | 6.82 | 0.06 | 5.13 | 302.6 |

Figure 3 Differences in the topology of Notch and Hedgehog embedded signaling pathways. The Notch (a) and Hedgehog (b) networks were constructed from HPRD data and the corresponding statistics are listed (c). Blue nodes represent the core components of pathways and yellow nodes represent their interacting proteins.

A)

| Pathway | KEGG | BioCarta |
|---|---|---|
| MAPK | 272 | 80 |
| TGF-β | 85 | 17 |
| Notch | 46 | 6 |
| Wnt | 148 | 25 |
| Hedgehog | 54 | 11 |
| mTOR | 50 | 22 |
| TLR | 92 | 36 |
| Jak-STAT | 154 | 5 |
| VEGF | 75 | 15 |

B)

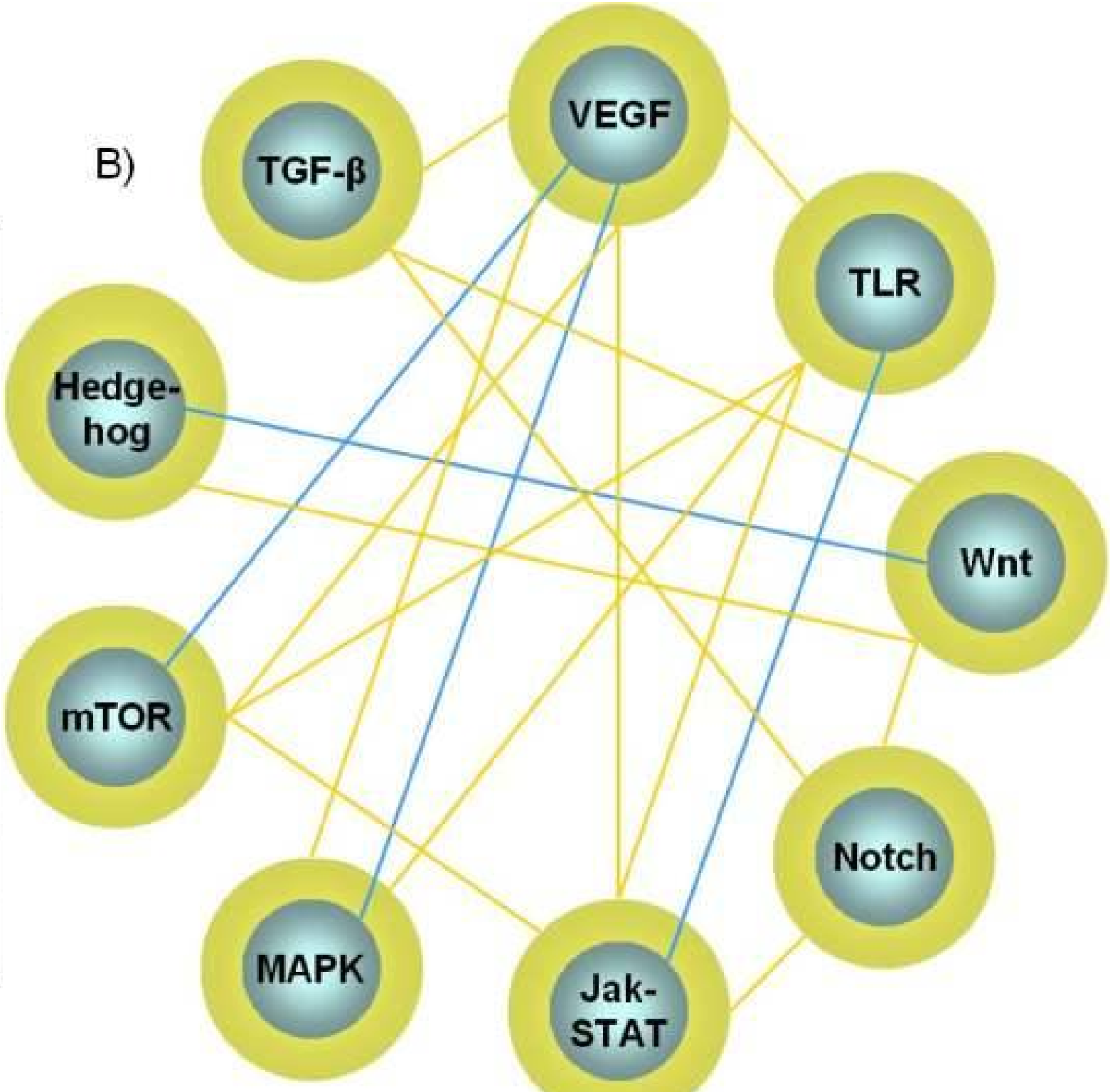

Figure 4 Overlaps between core and extended embedded pathways. (a) Discrepancies between the number of core components in KEGG and BioCarta are listed. (b) Blue nodes represent the core pathways, blue edges connect two core pathways that have significant overlaps (p-value ≤ 0.01). The yellow nodes and edges represent the extended pathways and the significant overlaps between extended pathways.

Figure 5. Biochemical pathways redrawn using the edge ontology defined in this paper. (a) Notch signaling pathway; (b) Citric acid Cycle; (c) Caspase cascade; (d) Jak-STAT signaling pathway)

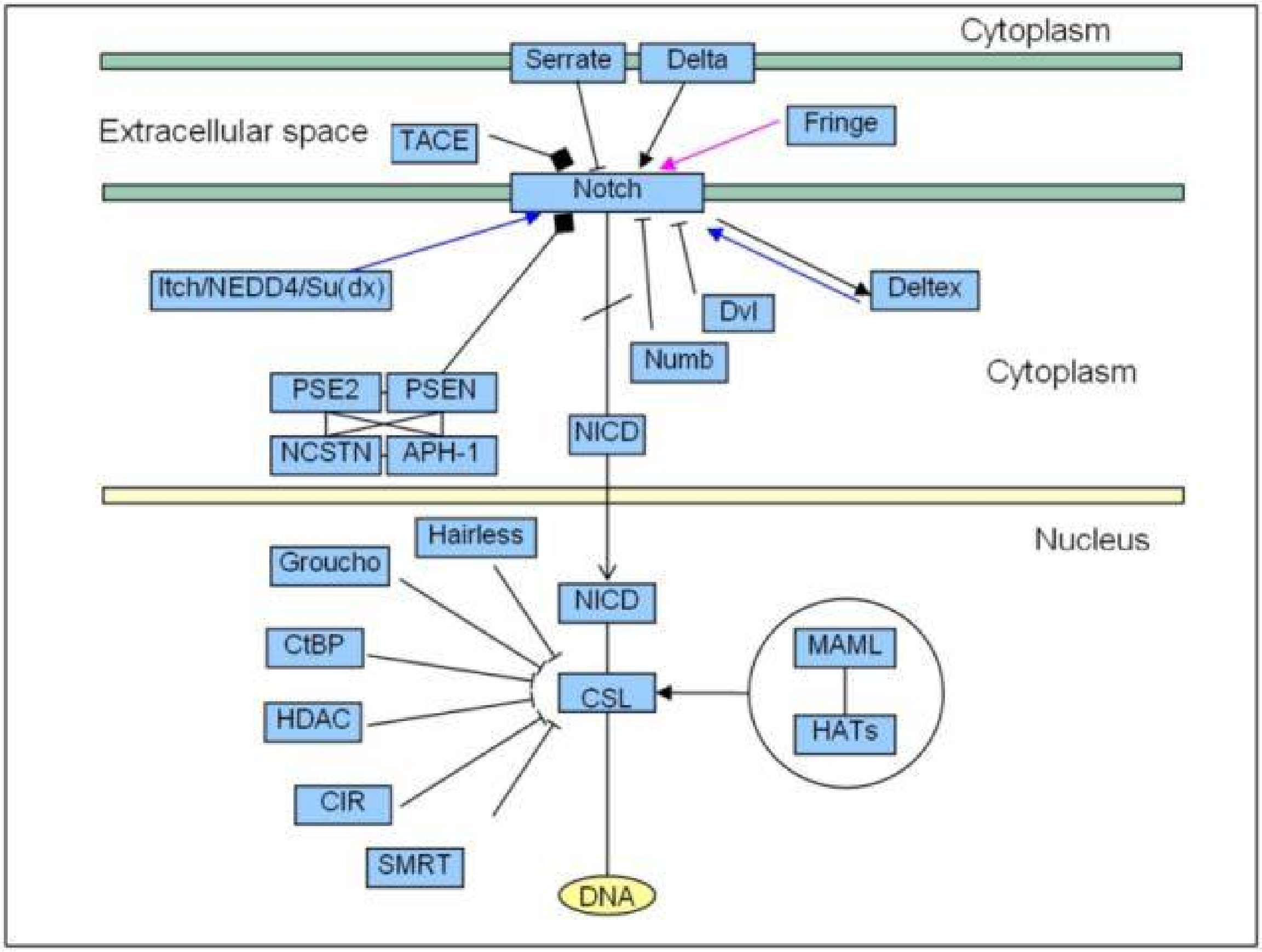

A).

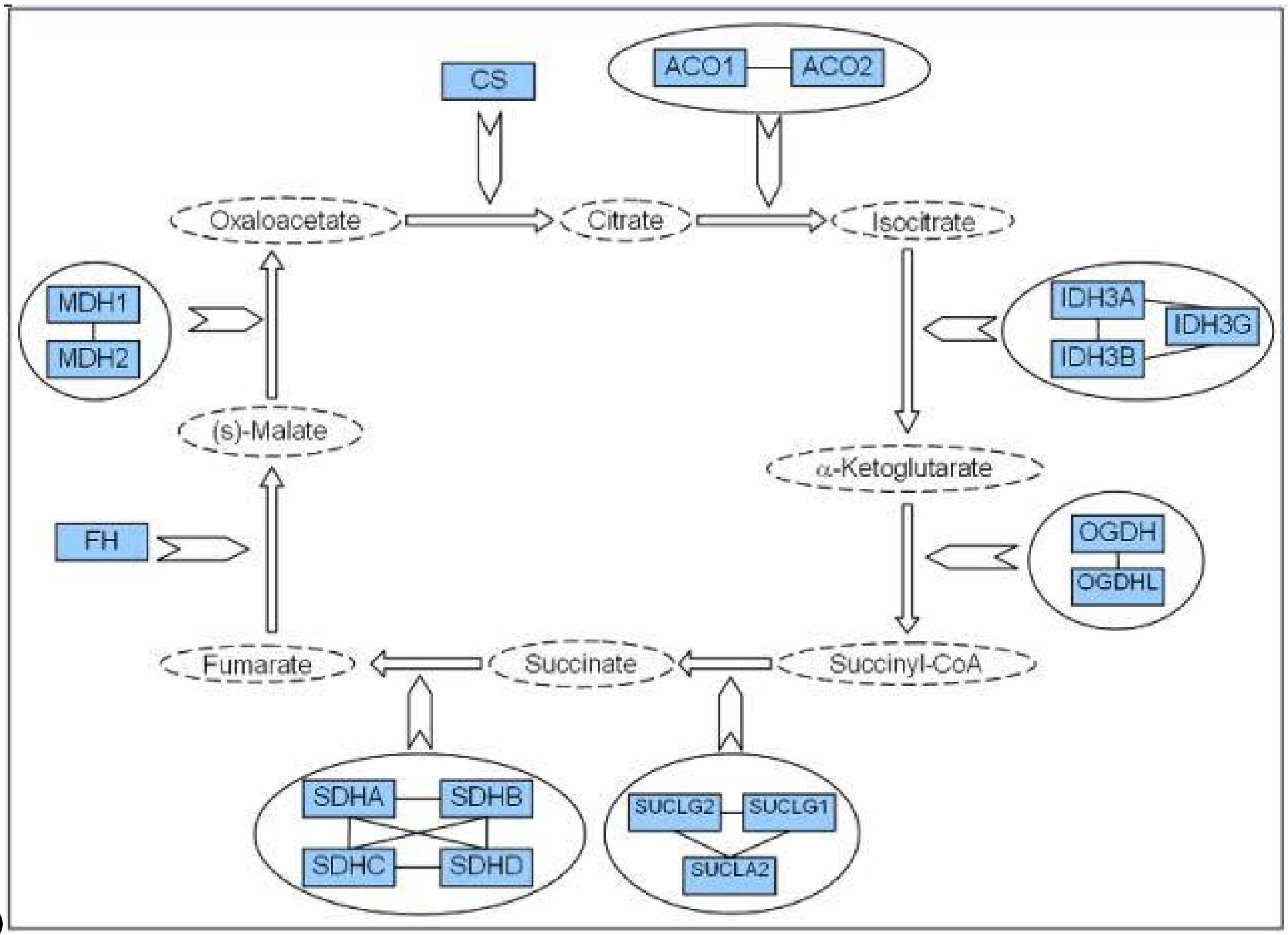

B)

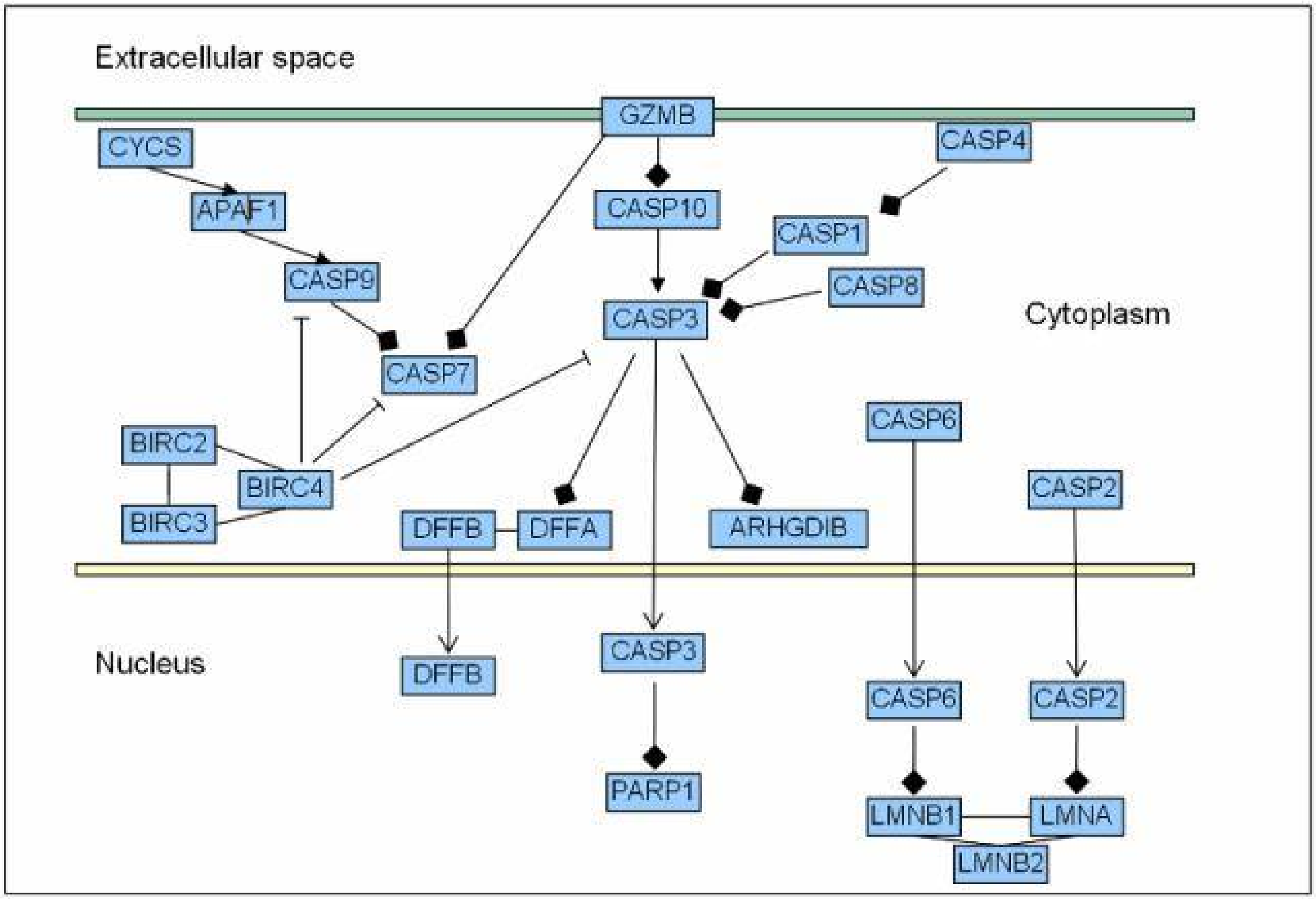

C)

D) 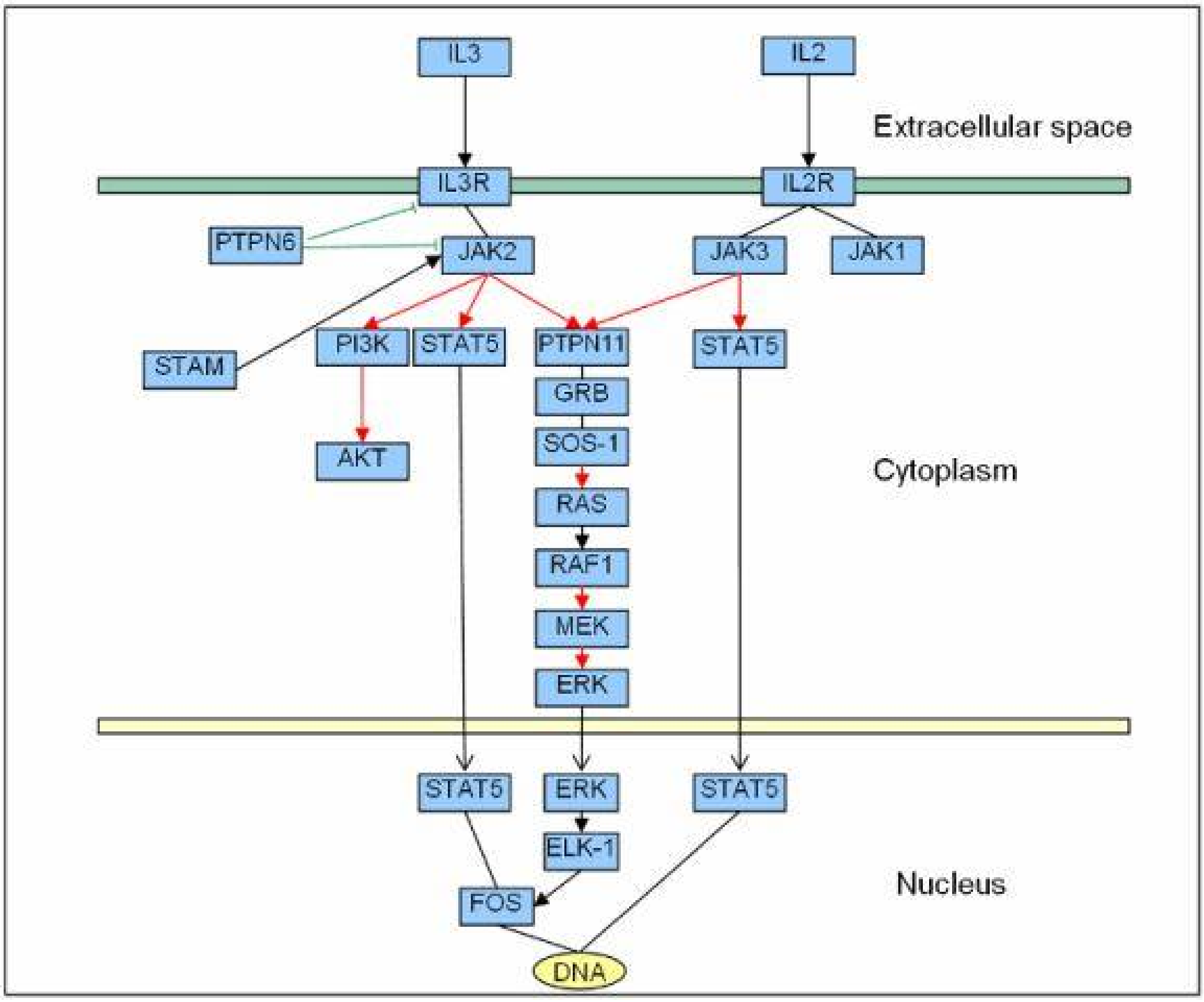

Table 1. A Simple Edge Ontology.

| | Direction (Level I) | Type (Level II) | Sub-type (Level III) | Specification (Level IV) |
|---|---|---|---|---|
| Interaction | Directed | Tagging of proteins[1] | Phosphorylation | serine[2] ser |
| | | | | tyrosine[2] tyr |
| | | | | other[2] aa |
| | | | Dephosphorylation | |
| | | | Ubiquitination | |
| | | | Glycosylation | N-linked N |
| | | | | O-linked O |
| | | | Methylation | |
| | | Cleavage of proteins | | |
| | | Translocation | Diffusion | |
| | | | Active transport | |
| | | Conformational change | | |
| | | Chemical reaction[3] | | |
| | | Catalysis | | |
| | | unknown/other | | |
| | Undirected | Binding [4] | | |
| | | Complex Assoc.[4]<br>Binding or Assoc.[4] | | |
| | | Dissociation | | |
| | | Co-expression | | |
| | | unknown/other | | |

[1] describes "inhibition" (where applicable)

[2] ser describes serine inhibition. A similarly annotated symbol can be used for tyrosine inhibition, etc.

[3] Double arrow describes reversible chemical reactions

[4] "Binding" describes direct physical interaction. "Association" describes two proteins that are linked in the same complex but do not directly physically interact. "Binding or association" describes the common scenario that arises in TAP-tagging experiments when one does not know the specific type of interaction.

*Supplemental Materials*

*Lu et al.* Comparing Classical Pathways and Modern Networks: Towards the Development of an Edge Ontology

## Relating network properties in embedded pathways

We first investigate whether commonly-used network measures are able to differentiate various types of pathways. Hierarchical clustering is performed for 66 extended embedded pathways mapped by corresponding pathways in KEGG (Table S5), based on six key network topological quantities: average degree (K), clustering coefficients (C), eccentricity (E), characteristic path length (L), diameter (D) and betweenness (B). The definition and biological relevance of these topological quantities are provided in Table S2. These embedded pathways can be divided into five categories by KEGG: Metabolism (M), Genetic information processing (GIP), Environmental information processing (EIP), Cellular processes (CP) and Human diseases (DIS) (Table S7).

Three main clusters are identified after the hierarchical clustering (Figure S2 and Table S5). The first and the third clusters show an enrichment of cellular processes and metabolism pathways, respectively. The second cluster is more heterogeneous and includes different types of pathways. Considering signaling and metabolic pathways only, we have 11 out of 17 (65%) signaling pathways in the first cluster, and 15 out of 17 (88%) metabolic pathways in the third cluster. Comparisons between the metabolic and signaling pathways reveal significant differences in average degree (K), eccentricity (E), and betweenness (B) (Figure S1). While these results seem to indicate the efficacy of employing network analysis to characterize pathways, we may have also captured the differences in mechanisms among these diverse types of pathways.

Figure S1. Network statistics of metabolic networks and signaling networks. The p-values of the differences are shown on top of each column pairs. They are computed by means of the Mann-Whitney test, which is the “non-parametric” version of the t-test. The comparison included 17 signaling networks and 17 metabolic networks.

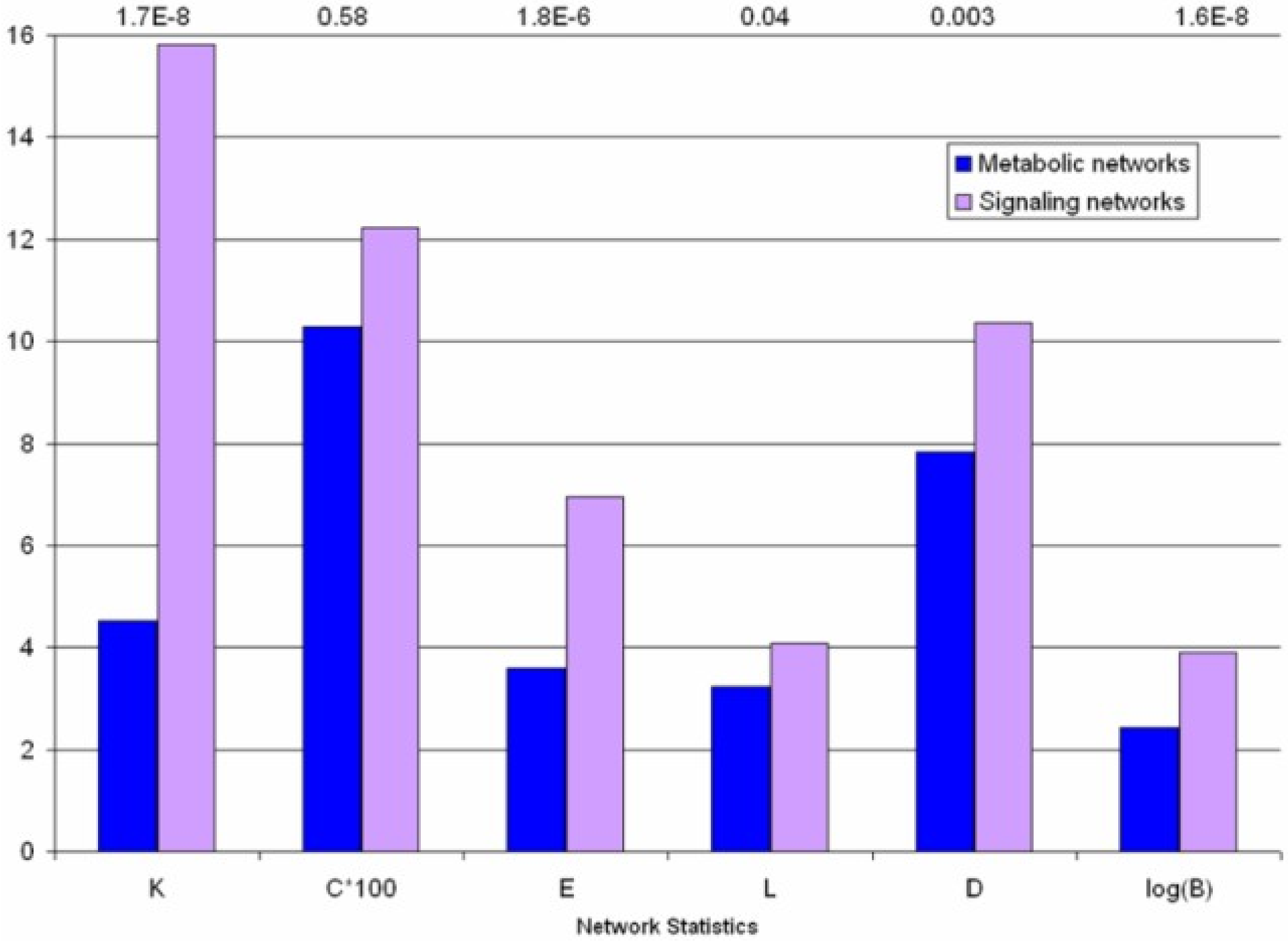

Figure S2. Clustering of 66 HTP networks based on five network statistics. On the left, we have the dendrogram representing the cluster. Each pathway category is identified by a color. The categories are those used in KEGG, namely, metabolism (red), genetic information processing (yellow), environmental information processing (light blue), cellular processes (blue), and human diseases (gray). The heat map provides a color representation of the network measurements. On the right of the heat map, the pathways are reported. The hierarchical clustering is computed as complete linkage with Euclidean distance.

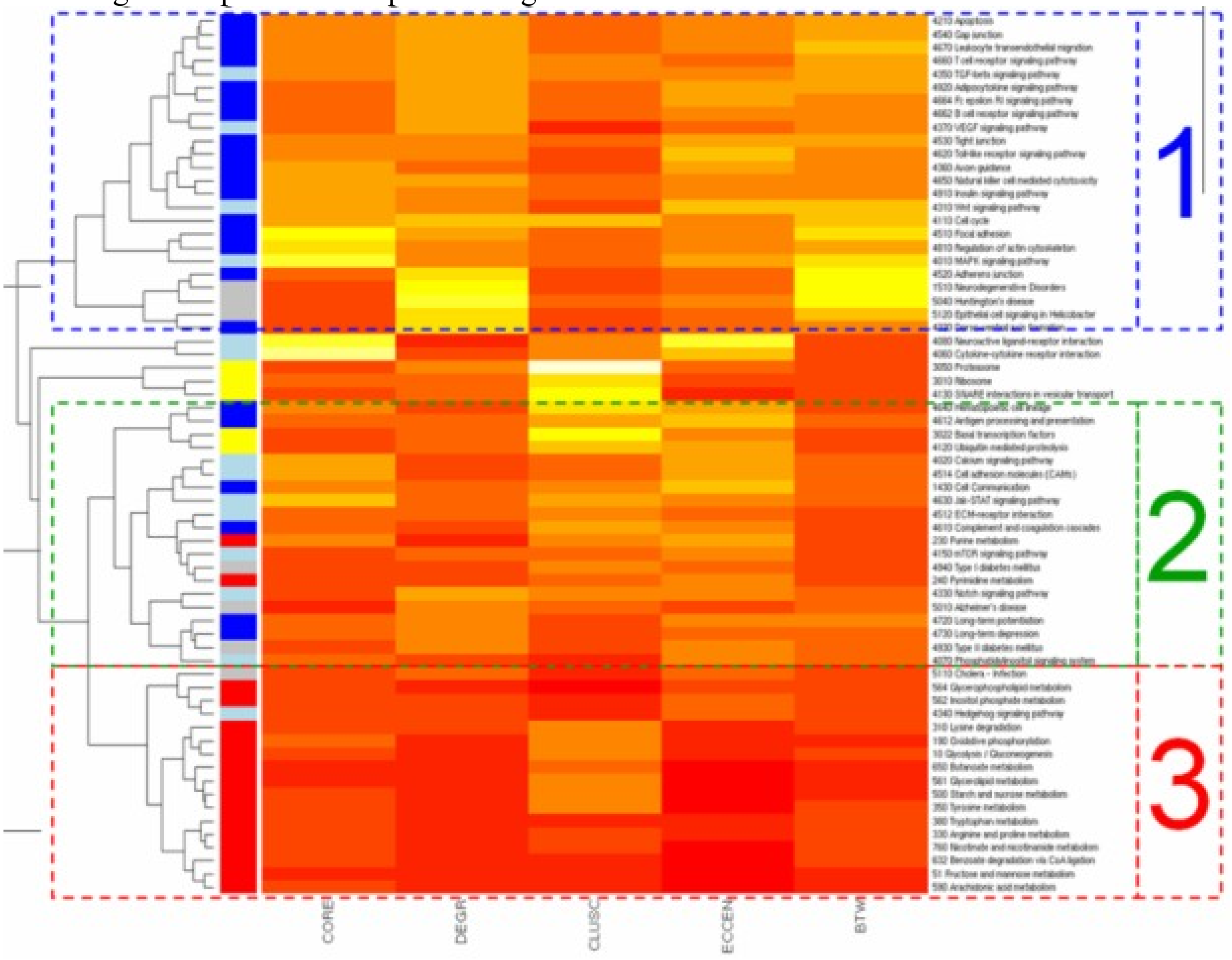

Figure S3. Correlations between network statistics.

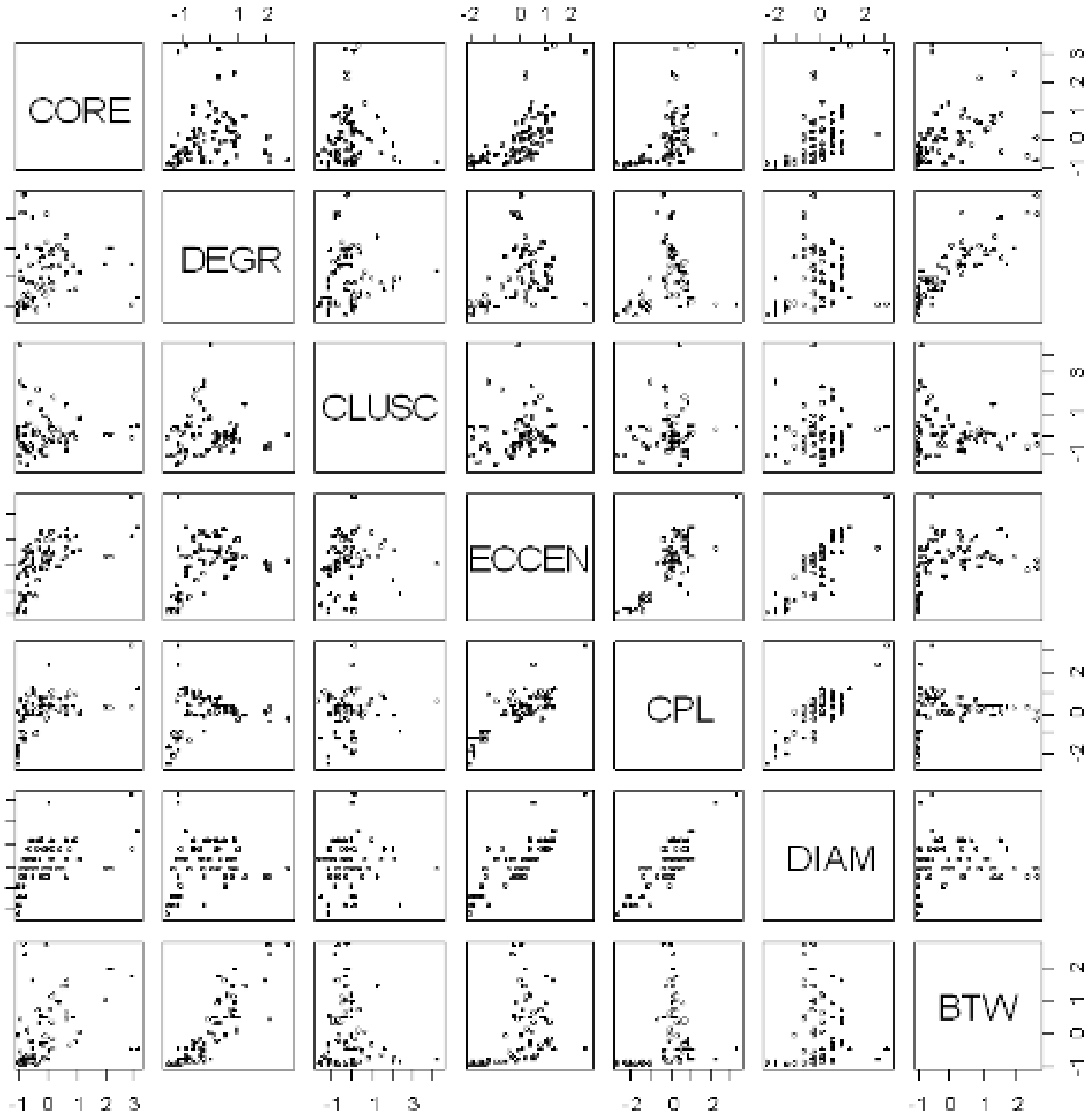

Table S1. Discrepancy between the number of core components in KEGG and BioCarta.

| **Pathway** | **KEGG ID** | **Number** | **BioCarta ID** | **Number** |
|---|---|---|---|---|
| MARK | Hsa04010 | 272 | H_mapkPathway | 81 |
| TGF-beta | Has04350 | 85 | H_tgfbetaPathway | 18 |
| Notch | Hsa04330 | 46 | H_notchPathway | 6 |
| Wnt | Hsa04310 | 148 | H_wntPathway | 20 |
| Hedgehog | Hsa04340 | 54 | H_sshPathway | 11 |
| mTOR | Hsa04150 | 50 | H_mTORPathway | 23 |
| Toll-like receptor | Hsa04620 | 92 | H_tollPathway | 37 |
| Jak-Stat | Hsa04630 | 154 | H_stat3Pathway | 6 |
| VEGF | Hsa04370 | 75 | H_vegfPathway | 15 |

Table S2. Summary of Topological Statistics of a Network.

| Network Statistics | Symbol | Definition |
|---|---|---|
| Average Degree | K | The degree of a node is the number of edges incident on it. The average degree of the whole network is the average of the degrees of all individual nodes. |
| Betweenness | B | The betweenness of a node n is defined as the number of node pairs ($n_1$, $n_2$) where the shortest path from $n_1$ to $n_2$ passes through n. |
| Characteristic Path Length | L | The graph-theoretical distance between two nodes is the minimum number of edges that is necessary to traverse from one node to the other. The characteristic path length of a network is the average of these minimum distances. It gives a measure of how closely nodes are connected within the network. |
| Eccentricity | E | The shortest path length, i.e., the eccentricity, of a node n is the length of its maximum shortest paths. |
| Clustering Coefficient | C | The ratio of the number of existing edges between a node's neighbors and the maximum possible number of edges between them (similar to an odds ratio). The clustering coefficient of a network is the average of all its individual clustering coefficients. This statistic can be used to determine the completeness of the network. |
| Diameter | D | The diameter of a network is the longest graph-theoretical distance between any two nodes in the graph. |

Table S3. Network topological measures of human signaling pathways after mapping onto the interaction network.

| Category | Node # | Edge # | Conn. Comp. | Core Node | Ext. Node | K | C | E | B |
|---|---|---|---|---|---|---|---|---|---|
| MAPK | 1759 | 3163 | 10 | 226 | 1533 | 15.85 | 0.12 | 8.11 | 16249.1 |
| mTOR | 370 | 466 | 10 | 46 | 324 | 10.72 | 0.13 | 6.35 | 2826.28 |
| Wnt | 1186 | 1749 | 9 | 110 | 1076 | 17.17 | 0.09 | 8.62 | 15160.8 |
| Notch | 459 | 683 | 3 | 36 | 423 | 20.47 | 0.17 | 6.64 | 6202.85 |
| Hedgehog | 212 | 238 | 9 | 38 | 174 | 6.82 | 0.06 | 5.13 | 302.6 |
| TGF-β | 781 | 1275 | 5 | 77 | 704 | 18.6 | 0.18 | 6.31 | 9662.59 |
| VEGF | 726 | 1154 | 6 | 62 | 664 | 19.53 | 0.06 | 5.65 | 8717.02 |
| TLR | 714 | 1087 | 3 | 79 | 635 | 14.78 | 0.10 | 8.38 | 7565.97 |
| Jak-Stat | 819 | 1498 | 3 | 131 | 688 | 13.47 | 0.20 | 6.94 | 5777.27 |

Table S4. Overlaps (crosstalk) and p-values for the overlaps between human signaling pathways. The number on the left of the slash is the percentage of overlaps between two core embedded pathways, and the number on the right is the percentage of overlaps between two extended embedded pathways. The two numbers in the parenthesis are the p-values for the corresponding overlaps.

| | **MAPK** | **mTOR** | **Wnt** | **Notch** | **Hedgehog** | **TGF-β** | **VEGF** | **TLR** | **Jak-Stat** |
|---|---|---|---|---|---|---|---|---|---|
| **MAPK** | | 0.1/0.3 (1/0.928) | 0.1/0.5 (1/1) | 0.0/0.2 (1/1) | 0.0/0.1 (1/1) | 0.1/0.4 (1/1) | 0.4/0.5 (0/0) | 0.2/0.5 (0.98/0) | 0.0/0.4 (1/1) |
| **mTOR** | | | 0.0/0.2 (1/1) | 0.0/0.1 (1/1) | 0.0/0.1 (1/0.996) | 0.1/0.2 (0.93/0.822) | 0.3/0.5 (0/0) | 0.2/0.4 (0.14/0) | 0.2/0.4 (0.6/0) |
| **Wnt** | | | | 0.1/0.4 (0.91/0) | 0.3/0.2 (0/0) | 0.2/0.5 (0.98/0) | 0.2/0.3 (0.63/1) | 0.1/0.4 (1/0.982) | 0.0/0.3 (1/1) |
| **Notch** | | | | | 0.0/0.1 (1/0.952) | 0.0/0.3 (1/0) | 0.0/0.1 (1/1) | 0.0/0.2 (1/1) | 0.0/0.3 (1/0.008) |
| **Hedgehog** | | | | | | 0.1/0.1 (0.66/0.988) | 0.0/0.1 (1/1) | 0.0/0.1 (1/1) | 0.0/0.1 (1/1) |
| **TGF- β** | | | | | | | 0.0/0.3 (1/0) | 0.0/0.3 (1/1) | 0.0/0.3 (1/0.816) |
| **VEGF** | | | | | | | | 0.2/0.4 (0.03/0) | 0.1/0.4 (1/0) |
| **TLR** | | | | | | | | | 0.2/0.4 (0.01/0) |
| **Jak-Stat** | | | | | | | | | |

Table S5. Sixty-six embedded pathways. M–metabolism; GIP–Genetic information processing; EIP–Environmental information processing; CP–Cellular processes; DIS–Human diseases.

| **Cluster 1** | | **Cluster 2** | | **Cluster 3** | | **Unclassified** | |
|---|---|---|---|---|---|---|---|
| Apoptosis | CP | Hematopoietic cell lineage | CP | Cholera - Infection | DIS | Neuroactive ligand-receptor interaction | EIP |
| Gap junction | CP | Antigen processing and presentation | CP | Glycerophospholipid metabolism | M | Cytokine-cytokine receptor interaction | EIP |
| Leukocyte transendothelial migration | CP | Basal transcription factors | GIP | Inositol phosphate metabolism | M | Proteasome | GIP |
| T cell receptor signaling pathway | CP | Ubiquitin mediated proteolysis | GIP | Hedgehog signaling pathway | EIP | Ribosome | GIP |
| TGF-beta signaling pathway | EIP | Calcium signaling pathway | EIP | Lysine degradation | M | SNARE interactions in vesicular transport | GIP |
| Adipocytokine signaling pathway | CP | Cell adhesion molecules (CAMs) | EIP | Oxidative phosphorylation | M | | |
| Fc epsilon RI signaling pathway | CP | Cell Communication | CP | Glycolysis / Gluconeogenesis | M | | |
| B cell receptor signaling pathway | CP | Jak-STAT signaling pathway | EIP | Butanoate metabolism | M | | |
| VEGF signaling pathway | EIP | ECM-receptor interaction | EIP | Glycerolipid metabolism | M | | |
| Tight junction | CP | Complement and coagulation cascades | CP | Starch and sucrose metabolism | M | | |
| Toll-like receptor signaling pathway | CP | Purine metabolism | M | Tyrosine metabolism | M | | |
| Axon guidance | CP | mTOR signaling pathway | EIP | Tryptophan metabolism | M | | |
| Natural killer cell mediated cytotoxicity | CP | Type I diabetes mellitus | DIS | Arginine and proline metabolism | M | | |
| Insulin signaling pathway | CP | Pyrimidine metabolism | M | Nicotinate and nicotinamide metabolism | M | | |
| Wnt signaling pathway | EIP | Notch signaling pathway | EIP | Benzoate degradation via CoA ligation | M | | |
| Cell cycle | CP | Alzheimer's disease | DIS | Fructose and mannose metabolism | M | | |
| Focal adhesion | CP | Long-term potentiation | CP | Arachidonic acid metabolism | M | | |
| Regulation of actin cytoskeleton | CP | Long-term depression | CP | | | | |
| MAPK signaling pathway | EIP | Type II diabetes mellitus | DIS | | | | |
| Adherens junction | CP | Phosphatidylinositol signaling system | EIP | | | | |
| Neurodegenerative Disorders | DIS | | | | | | |
| Huntington's disease | DIS | | | | | | |
| Epithelial cell signaling in H. pylori infection | DIS | | | | | | |
| Dorso-ventral axis formation | CP | | | | | | |

Table S6. Correlations between network statistics

| | CORE | DEGR | CLUSC | ECCEN | CPL | DIAM | BWN |
|---|---|---|---|---|---|---|---|
| **CORE** | 1.000 | | | | | | |
| **DEGR** | 0.111 | 1.000 | | | | | |
| **CLUSC** | 0.019 | -0.010 | 1.000 | | | | |
| **ECCEN** | 0.663 | 0.367 | 0.134 | 1.000 | | | |
| **CPL** | 0.529 | 0.168 | 0.076 | 0.819 | 1.000 | | |
| **DIAM** | 0.566 | 0.052 | 0.043 | 0.826 | 0.851 | 1.000 | |
| **BTW** | 0.372 | 0.879 | -0.174 | 0.399 | 0.198 | 0.116 | 1.000 |

Table S7. Summary of the distribution of the categories in each cluster by hierarchical clustering.

| **Cluster** | **Metabolism (M)** | **Genetic Information Processing (GIP)** | **Environmental Information Processing (EIP)** | **Cellular Processes (CP)** | **Human Diseases (DIS)** | **Total** |
|---|---|---|---|---|---|---|
| 1 | 0 | 0 | 4 | 17 (6E-6) | 3 | 24 |
| 2 | 2 | 2 | 7 (0.07) | 6 (0.79) | 3 | 20 |
| 3 | 15 (6E-11) | 0 | 1 | 0 | 1 | 17 |
| Unclassified | 0 | 3 | 2 | 0 | 0 | 5 |
| Total | 17 | 5 | 14 | 23 | 7 | 66 |